\documentstyle[12pt]{article}
\begin{document}
\rightline{NKU-SF2-2011}
\bigskip
\begin{center}
{\Large\bf New Charged Dilaton Solutions
 in 2+1 Dimensions and Solutions with Cylindrical Symmetry in 3+1 Dimensions

}

\end{center}
\hspace{0.4cm}
\begin{center}
{Sharmanthie Fernando \footnote{fernando@nku.edu}}\\
{\small\it Department of Physics \& Geology}\\
{\small\it Northern Kentucky University}\\
{\small\it Highland Heights}\\
{\small\it Kentucky 41099}\\
{\small\it U.S.A.}\\

\end{center}

\begin{center}
{\bf Abstract}
\end{center}

\hspace{0.7cm}{\small 

We report a new 
family of 
solutions to 
Einstein-Maxwell-dilaton 
gravity in 2+1 dimensions  and Einstein-Maxwell gravity with cylindrical
symmetry in 3+1 dimensions.
A set of static charged solutions
in 2+1 dimensions are obtained by a
compactification 
of charged solutions in 3+1 dimensions
with cylindrical symmetry.
These solutions contain naked singularities for
certain values of the parameters considered.
New rotating charged solutions in
2+1 dimensions and 3+1 dimensions
are generated treating the static
charged solutions as  seed metrics
and performing $SL(2;R)$ transformations.}
\newpage

\section{Introduction}
Three dimensional gravity has provided us with many 
important clues about higher dimensional physics.
String theory, which seems to be the best candidate 
available for a consistent theory of quantum gravity,
requires studies of low dimensional
effective string  actions.
In this respect, dilaton gravity in 2+1 dimensions 
deserve further attention since it arises from a low
energy string effective theory. 

In one of the earlier works, Shiraishi\cite{shiraishi}
found a family of static multi centered solutions for
Einstein-Maxwell-dilaton
gravity. Park and Kim \cite{park}
constructed general static axially symmetric
solutions to the same model 
by dimensional reducing to two dimensions.
In recent times, lot of attention has been given
to gravity in 2+1 dimensions
with a negative
cosmological constant due to the existence of
black hole solutions \cite{banados}. Modification
of this black hole
with a dilaton and Maxwell's fields 
have lead to many interesting results.
Static charged black holes
and spinning
black holes 
in anti-de Sitter space by Chan and Mann\cite{chan2}\cite{chan1},
spinning solutions with self dual electromagnetic fields
by Fernando\cite{fer},
black holes in a generalized dilaton gravity with
a Brans-Dicke type parameter by S\'{a} {\it et.al.}\cite{lemo},
magnetic solutions by Koikawa {\it et.al.}\cite{koi} are
some of the work related to dilaton gravity
in anti-de Sitter space.

In this paper we present an interesting class of
dilaton solutions arising from four
dimensional gravity. This is achieved by a
compactification of 
electrically and magnetically charged
solutions with
cylindrical symmetry in four dimensions.
These new set of solutions are different from
the ones presented above due its
direct relation to four dimensions.
These solutions are important 
since it provide us with a
clear understanding of 
how  4D gravity and 3D gravity are related to each other.
Furthermore, we have generated rotating charged solutions
to Einstein-Maxwell-dilaton gravity via ``T-duality''
and treating the  compactified solutions as the seed metrics.
Also, embedding these 2+1 dimensional
solutions in 3+1 dimensions leads
to new rotating charged solutions with cylindrical symmetry. Such a compactification
from an uncharged cylindrical solution to 2+1 dimensions is reported in Fernando \cite{fer2}.

We have structured the paper as follows.
In section 2 we will give a brief introduction
to four dimensional charged solutions with
cylindrical symmetry and compactify them to obtain dilaton solutions
in three dimensions. 
In section 3 and 4 we will study static
charged solutions. In section 5, new rotating charged solutions
in 2+1 dimensions 
are generated. In section 6, new rotating charged solutions
in 3+1 dimensions with cylindrical symmetry
are presented. Finally we will conclude.

\section{Compactification
of Charged Solutions in Four Dimensions
}

Cylindrical symmetrical space-times
in four dimensions consists of  isometries generated
by two commuting space-like Killing vectors.
If the solutions are stationary, then the space-time
admit another Killing vector along the time axis.
General stationary cylindrical symmetric
line element with three Killing vectors 
$\partial_t, \partial_z,$ and $\partial_{\varphi}$ can be written as,
\begin{equation}
ds^2 = e^{-2U} \left( e^{2K} ( dr^2 + dz^2 ) + r^2 d \varphi^2 \right)
-e^{2U} ( dt+ A d \varphi)^2
\end{equation}
where $U, K$ and $A$ are functions of $r$ only. The above metric
is a solution to the Einstein-Maxwell action in four dimensions given by,
\begin{equation}
S =    \int d^4x \sqrt{-G} ( R - F_{\alpha \beta} F^{\alpha \beta})
\end{equation}
The purpose of this section is to  dimensionally reduce
four dimensional Einstein gravity 
to three dimensions
to obtain dilaton gravity. If we pick the cylindrical 
solutions  given above
and treat the $z$ coordinate to
be compact
for the purposes of compactification,
the reduction will yield a theory
in three dimensions with gravity, dilaton field $\phi$
and the gauge field $A_{\mu}$.

Now, to perform compactification along the
$z$ direction, let us 
rewrite the above metric in four dimensions as
follows:
\begin{equation}
 ds^2_{3+1} = G_{\mu \nu} dx^{\mu} dx^{\nu}=
 {g}_{ab} dx^a dx^b + e^{-4\phi}dz^2
 \end{equation}
 with
 \begin{equation}
 e^{-4 \phi} = G_{zz} 
 \end{equation}
 Here $ (\mu, \nu = 0,1,2,3)$ 
 and $(a,b = 0,1,2 )$ are four and 
 three dimensional indices respectively.
 The  dimensionally reduced action in three dimensions
is given by,
 \begin{equation}
 S_{string} = \int d^3x  \sqrt{-g^S} e^{-2 \phi} (R^S -  F_{\alpha \beta} F^{\alpha \beta})
 \end{equation}
 Here, $g^S_{ab}$ corresponds to $g_{ab}$ of the four dimensional
 metric and the metric can be treated to be in the ``string frame''.
 One can perform a conformal transformation
 to bring the metric to Einstein frame as follows,
 \begin{equation}
 e^{-4 \phi}g^S_{\mu \nu} = g^E_{\mu \nu}
 \end{equation}
 This transformation  will lead to the following action and the 
 corresponding field equations,
 \begin{equation}
 S_{Einstein} = \int d^3x \sqrt{-g} ( R - 
 8 \nabla_{\mu} \phi \nabla^{\mu} \phi - e^{-4\phi}F^2)
 \end{equation}
 \begin{equation}
R_{\mu \nu} = 8\nabla_{\mu} \phi \nabla_{\nu} \phi + e^{-4 \phi} ( -g_{\mu \nu} F^2
+ 2 F_{\mu}^{\theta} F_{\nu \theta})
\end{equation}
\begin{equation}
8\nabla_{\mu}\nabla^{\mu} \phi + 2 e^{-4 \phi} F^2 = 0
\end{equation}
\begin{equation}
\nabla_{\mu} ( e^{-4\phi} F^{\mu \nu}) = 0
\end{equation}

\section{Static Electrically Charged Dilaton Solutions}
First we will consider the static cylindrical solutions  of Einstein-Maxwell
gravity corresponding to a radial
electric field caused by an axial charge distribution
along the z-axis is given as follows:
\begin{equation}
ds^2 = r^{2m^2}L^2 ( dr^2 + dz^2 ) + r^2 L^{2}d \varphi^2  )
-L^{-2}dt^2
\end{equation}
The function $L$ is given by,
\begin{equation}
L = c_1r^m + c_2 r^{-m}
\end{equation}
with $c_1, c_2$ and $m$ being real
constants. This metric is a solution of the action in eq.(2).
The electric field is given by 
\begin{equation}
F_{rt} = \frac {-Q}{L^2r}
\end{equation}
with,
\begin{equation}
-4c_1 c_2 m^2 =   Q^2
\end{equation}
Notice that $c_1$ and $c_2$ need to have opposite signs for the charge $Q$ 
to be real. In the following discussion we will consider $c_1 < 0$
and $c_2 > 0 $. For $c_1 =0$ and $c_2 =1$,
the above metric corresponds to Levi-Civita vacuum
static solution\cite{kramer}.
The Mukherji \cite{muck} solution describing the
gravitational field of a charged line-mass is contained in 
eq.(11). Now, by starting from the
metric in eq.(11), one can obtain solutions to three dimensional space-time
after compactification as,
\begin{equation}
ds_{Ein}^2 = r^{4m^2}L^4 dr^2  + r^{2+2m^2}L^4 d \varphi^2 )
- r^{2m^2}dt^2
\end{equation}
The space-time in 2+1 dimensions consists of a dilaton field given by,
\begin{equation}
\phi = -\frac{1}{2} ln \left(r^{m^2} ( c_1 r^m + c_2 r^{-m})\right)
\end{equation}
and a electric potential given by,
\begin{equation}
A_t = \frac{Q}{2c_1m} \frac{r^{-m}}{(c_2r^{-m} + c_1 r^{m})}
\end{equation}
The Ricci Scalar $Rs = R_{\mu \nu}g^{\mu \nu}$,  Kretschmann scalar
$K = R_{\mu \nu \gamma \beta} R^{\mu \nu \gamma \beta}$ and  
$F^2= F_{\mu \nu} F^{\mu \nu}$ 
are computed for the above metric as follows:
\begin{equation}
Rs = \frac{( 2m^2 r^{-2 + 4m -4m^2})}{(1+c_1r^{2m})^6}   \left\{ ( 1 -2m + m^2) + 2r^{2m} c_1 ( m^2 -3) +
r^{4m} c_1^2 ( 1 + 2m+ m^2 ) \right\}
\end{equation}
\newpage
$$
K = \frac{(4m^4 r^{-4+8m-8m^2})}{(1+c_1r^{2m})^12} \{ 3 - 12m+18m^2 - 12m^3 + 3m^4 +$$
$$r^{2m}( -4c_1 + 8c_1m + 8c_1 m^2 -24c_1m^3 + 12c_1m^4)
+ r^{4m}(50c_1^2 - 20c_1^2m^2 + 18c_1^2m^4)$$
$$ r^{4m}(50c_1^2 - 20c_1^2 m^2 + 18c_1^2 m^4) + r^{6m} ( -4c_1^3 -8c_1^3m + 8c_1^3m^22)
$$
\begin{equation}
r^{8m}(3c_1^4 + 12c_1^4m + 18m^2c_1^4 + 12c_1^4m^3 + 3c_1^4m^4) \}
\end{equation}
\begin{equation}
F^2 =  \frac{-2Q^2}{( c_2 + c_1 r^{2m})^8} r^{( -2 + 8m - 6m^2)}
\end{equation}

\subsection{Case 1:  $c_1= 0$}
In this case the metric becomes,
\begin{equation}
ds^2 = r^{4m^2 -4m}c_2^4 dr^2 + r^{2+2m^2 -4m} c_2^4 d\varphi^2 - r^{2m^2}dt^2
\end{equation}
and the charge $Q = 0$.
For $m=0$ the above metric can be
shown to be
flat when the coefficient $c_2$ is absorbed
into the definition of $r$ and $t$. For $m=1$,
the resulting metric can be shown to be flat by a
coordinate transformation $ \varphi \rightarrow it$ 
and $ t \rightarrow -i\varphi$.
In fact one can observe
that the  scalar curvature $Rs = 0$ for $m=0, 1$ and
the dilaton,
\begin{equation}
\phi = -\frac{1}{2} ( m^2 -m) ln(r)
\end{equation}
is zero for $m=0,1$ as expected.
But
for $m \neq 0,1$ 
there will be a curvature singularity at $r =0$ (given $m \geq 0$) 
and the dilaton will be non-zero.

\subsection{Case 2: $c_1 \neq 0$  and  $c_2 =1$}
When $c_1 \neq 0$ and $m=0,1$, the charge $Q=0$ and the dilaton
$\phi =0$ leading to a flat
space-time.
For $c_1 \neq 0$ and  $m \neq 0,1$, the electric charge is non-zero.
Furthermore there are two curvature singularities
in the metric,
one at $r_{s1}= (-1/c_1)^{1/2m}$ and the other at $r_{s2}=0$.
However,  since $c_1 < 0$, the singularity at $r_{s1} > 0$
results in a  naked singularity.

\section{Static Magnetically Charged  Dilaton Solutions}

Here, we will consider static 
cylindrical symmetric solutions
of Einstein-Maxwell gravity corresponding to a magnetic field
along the z-axis. Physically it could
correspond to the field interior to 
a solenoid current whose axis is the z-axis\cite{kramer}. The metric for such a space-time is given by,
\begin{equation}
ds^2 = r^{2m^2}L^2 ( dr^2 - dt^2 ) +  L^{-2}d \varphi^2 )
+ r^2L^{2}dz^2
\end{equation}
The function $L$ is given by,
\begin{equation}
L = c_1r^m + c_2 r^{-m}
\end{equation}
with $c_1, c_2$ and $m$ being real
constants.
The magnetic field is given by 
\begin{equation}
F_{r\varphi} = \frac {Q}{L^2r}
\end{equation}
with,
\begin{equation}
4c_1 c_2 m^2 =   Q^2
\end{equation}
Note that for the charge $Q$ to be real,
both $c_1$ and $c_2$ should have the same sign
in contrast to the electric case.
For $m =1$, $c_1 = - \frac{B_0^2}{4}$ and $c_2 = -1$,
the above metric corresponds to,
\begin{equation}
ds^2 = \left( 1 + \frac{B_0^2 r^2}{4} \right)^2 ( dr^2 + dz^2 -dt^2 ) +  
\left( 1 + \frac{B_0^2 r^2}{4} \right)^{-2} d \varphi^2
\end{equation}
This is widely known as Melvin  magnetic universe (or ``flux tube'')
\cite{melvin} which represents a gravitational field generated by
a uniform magnetic field $B_0$ along the z-axis.

One can obtain magnetically charged solutions
to 2+1 dimensional Einstein-Maxwell-dilaton gravity by
compactifying the above metric in eq.(23) as follows:
\begin{equation}
ds_{Ein}^2 = r^{2+2m^2}L^4 dr^2  + r^{2} d \varphi^2
- r^{2+2m^2}L^4 dt^2
\end{equation}
with a dilaton,
\begin{equation}
\phi = -\frac{1}{2} ln \left(r ( c_1 r^m + c_2 r^{-m})\right)
\end{equation}
and a magnetic potential  given by,
\begin{equation}
A_{\varphi} = \frac{-Q}{2c_1m} \frac{r^{-m}}{(c_2r^{-m} + c_1 r^{m})}
\end{equation}
The Ricci scalar $Rs$, Kretschmann scalar $K$ and $F^2 = F_{\mu \nu}F^{\mu \nu}$
are computed for the above metric as follows:
\begin{equation}
Rs= \frac{r^{(-4 + 4m -2m^2)}} { ( c_2 + c_1 r^{2m})^6}
\left[ c_2^2 ( 1- m )^2 + 2 c_1 c_2 r^{2m}(1- 3m^2 )
+ c_1^2 r^{4m} ( 1 + m )^2 \right]
\end{equation}
$$
K = \frac{4 r^{ - 8 + 8 m -4m^2} }{ (c_2+c_1 r^{2m})^2 }
\{ 3c_2^4 ( 1 - m)^4 + 4r^{2m} c_1 c_2^3 ( 3 - 6m + 2m^2 + 2m^3 - m^4 ) 
$$
$$
+ r^{4m} (c_1c_2)^2 ( 18 - 20 m^2 + 50 m^4 )
+ 4r^{6m} c_1^3 c_2 ( 3 + 6m +  2m^2 - 2m^3 - m^4)
$$
\begin{equation}
+ 3r^{8m}c_1^4 ( 1+ m )^4 \}
\end{equation}
  \begin{equation}
F^2 = \frac{2 Q^2 r^{-6 -2m^2} } {( c_2 r^{-m} + c_1 r^{m} ) ^8 }
\end{equation}

\subsection{Case 1:  $c_1= 0$}
In this case the metric becomes,
\begin{equation}
ds^2 = r^{4m^2 -4m}c_2^4 dr^2 + r^{2+2m^2 -4m} c_2^4 d\varphi ^2 - r^{2m^2}dt^2
\end{equation}
and the charge $Q = 0$. Now, we will consider special values of $m$
to understand the structure of these space-times.
First, for $m=0$ the above metric is,
\begin{equation}
ds^2 = r^2 c_2^4 dr^2 + r^2 d \varphi ^2 - r^2 c_2 ^4 dt^2
\end{equation}
and the dilaton $\phi$ = $\frac{-1}{2} ln[rc_2]$. From  eq.(31,32), one can observe that
the scalar curvatures are non-zero for $m =0$. Hence in contrast to
the electric case, the space-time is not flat  for $m = 0$.
For $m =1$, the metric is given by,
\begin{equation}
ds^2 = c_2^4 dr^2 + r^2 d \varphi ^2 -  c_2 ^4 dt^2
\end{equation}
By redefining $r$ to be $r/c_2^2$ and $t$ to be $t c_2^2$,
the above metric becomes,
\begin{equation}
ds^2 = dr^2 + r^2c_2^{-4} d \varphi ^2 -  dt^2
\end{equation}
and the dilaton $\phi$ = $\frac{-1}{2} ln[c_2]$. The scalar curvatures $Rs$ and 
$K$ are zero. Hence, the space-time in eq.(37) corresponds to
a conical singularity similar to the one obtained by Deser et.al.\cite{deser}
and Gott\cite{gott}. The appearance of the
conical singularity can be understood as follows:
If we replace $\varphi$ with $\hat{\varphi} = \varphi/c_2^2$, 
then the metric eq.(37) corresponds to flat space-time locally. However,
the  former periodic coordinate $\varphi$
has the range
$ 0 \leq \varphi \leq 2 \pi$ and the new period coordinate
has the range
$ 0 \leq \hat{\varphi} \leq 2 \pi c_2^{-2}$.
Hence there is a deficit angle $D$ at the origin
due to the presence of a massive source
given by $D = 2\pi ( 1 - c_2 ^ {-2})$
leading to the conical space-time.
If $c_2 = 1$, then $D =0$ and the singularity vanishes.
This is expected since the dilaton vanishes for $c_2 = 1$
leading to  a vacuum.
For $m \neq 0,1$, there is a curvature singularity at $r=0$
and is not covered by a horizon.

\subsection{Case 2: $c_1 \neq 0$  and  $c_2 =1$}

In this case, if $m=0$ the electric charge is zero. However,
in contrast to the electric case,
the dilaton $\phi$ = -$\frac{1}{2} ln[r(c_1 + c_2)]$  is not a constant.
Observing the metric,
\begin{equation}
ds^2 = r^2 (c_1 + c_2)^4 ( -dt^2 + dr^2 ) + r^2 d \varphi ^2
\end{equation}
it is clear that the space-time is not flat. Furthermore,
there is a  curvature singularity at $r = 0$.
For $ m \neq 0$ the space-time is non-trivial  and corresponds to
a charged-dilatonic space-time.

 \section{Generating Rotating Charged Solutions in 2+1 Dimensions}

Treating the above static charged solutions as seed metrics,
one can generate rotating charged solutions in 2+1 dimensions.
Here, we adopt a solution generating technique developed by
Chan\cite{chanche}. Note that the static charged solutions presented in the previous
section have two commuting Killing vectors $\partial_t$ and 
$\partial _{\varphi}$. One can write a general metric with
such isometries  as,
\begin{equation}
ds^2 = h_{ij} dx^i dx^j + e^{\chi} P(r)^{-2}dr^2
\end{equation}
where
\begin{equation}
x^i = (\varphi,t); \hspace{1.0cm} 
\chi = ln|det h|; \hspace{1.0cm} \mbox{and} \hspace{1.0cm}   A = A_i dx^i
\end{equation}
Here, $A_r$ is choosen to be zero as a pure gauge quantity. Now,
substituting
the metric in eq.(39) in action  eq.(7),
and dividing it by the two-dimensional volume of the Killing orbits,
one arrives at,
\begin{equation}
S = \int dr  \left \{ P \left( \frac{\chi'^2}{8} + \frac{1}{4} Tr(H' H'^{-1})
- 8\phi'^2 - 2 e^{-4 \phi -\chi/2} Tr ( A'^{T} H^{-1} A') \right) \right \}
\end{equation}
Here,
\begin{equation}
H_{ij} = e^{-\chi /2} h_{ij} ;\hspace{1.0cm} H_{ij}^{-1} = e^{\chi/2}h^{ij}
\hspace{1.0cm} \mbox{and  $A$ is a column vector of  $A_i$.}
\end{equation}
The above action is invariant under $SL(2;R)$
transformations given by,
\begin{equation}
 H \rightarrow \Omega^T H \Omega, \hspace{1.0cm}  A \rightarrow \Omega^T A
 \hspace{1.0cm}\mbox{where $\Omega$ $\epsilon$ $SL(2;R)$}
 \end{equation}
 The above symmetry in the Lagrangian is similar to
 the ``target space'' symmetry arising in classical
 string effective action given by the 
 ``T-duality'' group $O(1,2;R)$ \cite{ortin}.
 Now, to perform the transformations in eq.(43), 
an arbitrary value of $\Omega$ can be
 parameterized as,
 \begin{equation}
 \Omega = e^{\gamma J_{+}} e^{\eta J_{1}} e^{\beta J_{-}}
 \end{equation}
 Here, $J_{a}$ are $SL(2;R)$ generators which has  matrix
  representation,
\begin{eqnarray}
J_{0} =  \left(\begin{array}{cc} 0 & 1 \\ -1 & 0
\end{array} \right) \;\;\;\; ;
J_{1} =  \left(\begin{array}{cc}  -1 & 0 \\ 0 & 1
\end{array} \right) \;\;\;\; ;
J_{2} = \left(\begin{array}{cc} 0 & 1 \\ 1 & 0
\end{array} \right)
\end{eqnarray}
and $J_{\pm} = (J_{2} \pm J_{0})/2$. Taking $e^{-\eta} = \alpha$
in the parameterization in eq.(44) will lead to the following $\Omega$: 
 \begin{eqnarray}
 \Omega = 
 \left(\begin{array}{cc} \alpha & \alpha \beta 
 \\ \beta \gamma & \alpha^{-1} + \alpha \beta \gamma \end{array} \right)
 \end{eqnarray}

\subsection{Rotating Electrically Charged Dilaton Solutions}

Considering the static charged metric in eq.(15) as the seed metric, one can 
apply the transformations given above to obtain the new metric as,
\begin{equation}
ds_{new}^2 = \tilde{h}_{ij}dx^i dx^j + r^{4m^2} L^4 dr^2
\end{equation}
where $\tilde{h_{ij}}$ are given by,
$$\tilde{h}_{tt} = -r^{2m^2} \alpha^2 + r^{2 + 2m^2} ( c_2 r^{-m} + c_1 r^{m} ) ^{4} \alpha^2 \gamma^2$$
$$\tilde{h}_{t\varphi} = -r^{2m^2} \alpha^2 \beta + 
r^{2 + 2m^2} ( c_2 r^{-m} + c_1 r^{m} ) ^{4} \alpha \gamma ( \alpha^{-1} + \alpha \beta \gamma)
$$
\begin{equation}
\tilde{h}_{\varphi \varphi} = -r^{2m^2} \alpha^2 \beta^2 + 
r^{2 + 2m^2} ( c_2 r^{-m} + c_1 r^{m} ) ^{4} ( \alpha^{-1} + \alpha \beta \gamma)
\end{equation}
The new potential is,
\begin{equation}
\tilde{A_i} dx^i = \frac{\alpha Q}{2 c_1 m} 
\frac{r^{-m}}{(c_2 r^{-m} + c_1 r^{m})} ( dt + \beta d\varphi)
\end{equation}
The dilaton remains the same under these transformations.

\subsection{Rotating Magnetically Charged Dilaton Solutions}

Considering the  magnetically 
charged static metric in eq.(28) as the seed metric, one can 
apply the transformations given above to obtain the new metric as,

\begin{equation}
ds^2_{new} = \tilde{h}_{ij}dx^i dx^j + r^{2+2m^2} L^4 dr^2
\end{equation}
where $\tilde{h_{ij}}$ are given by,
$$\tilde{h}_{tt} = r^{2} \alpha^2 \gamma^2 - 
r^{2 + 2m^2} ( c_2 r^{-m} + c_1 r^{m} ) ^{4} \alpha^2$$
$$\tilde{h}_{t\varphi} = 
r^{2} \alpha \gamma ( \alpha^{-1} + \alpha \beta \gamma)
-r^{2 + 2m^2} ( c_2 r^{-m} + c_1 r^{m} ) ^{4} \alpha^2 \beta$$
\begin{equation}
\tilde{h}_{\varphi \varphi} = r^{2}  
( \alpha^{-1} + \alpha \beta \gamma)^2
-r^{2 + 2m^2} ( c_2 r^{-m} + c_1 r^{m} ) ^{4} \alpha^2 
\beta^2
\end{equation}
The new potential is,
\begin{equation}
\tilde{A_i} dx^i = \frac{-\alpha \gamma Q}{2 c_1 m} 
\frac{r^{-m}}{(c_2 r^{-m} + c_1 r^{m})} \left( dt + ( \alpha^{-2} \gamma^{-1} 
+\beta) d\varphi \right)
\end{equation}
The dilaton remains the same under these transformations.

\section{Rotating Charged Solutions in 3+1 Dimensions with
Cylindrical Symmetry}

Solutions with cylindrical symmetry in 3+1
dimensions have interesting features. Cosmic strings,
which is one feature of gravitational collapse in cylindrical
symmetry are of particular interest since they are
considered as possible ``seeds'' for galaxy formation
and gravitational lenses. Furthermore, the
ease with which naked singularities are formed
in cylindrical symmetry needs further attention
due to the apparent failure of cosmic censorship. Unlike
the spherical gravitational collapse,
collapse of cylindrical systems are not well understood.
From this point of view, studying 
solutions with cylindrical symmetry is a worthy cause.
Here, we present a new class of rotating charged cylindrical
solutions. These are obtained by embedding the 
Einstein-Maxwell-dilaton solutions n 2+1 dimensions in 3+1
dimensions.

Now, we  shall describe how the 
above rotating charged solutions in 2+1 dimensions can be
embedded in 3+1 dimensions to obtain cylindrical solutions.
Suppose, the new  charged-dilaton solutions in 2+1 dimensions are given
by,
\begin{equation}
ds^2 = \tilde{h}_{ij}dx^i dx^j +  e^{\chi} P(r)^{-2} dr^2
\end{equation}
as represented in eq.(39), with a dilaton field $\phi$, then, the four dimensional
solution is given by,
\begin{equation}
ds^2 = \left(\tilde{h}_{ij}dx^i dx^j +  e^{\chi} P(r)^{-2} dr^2 \right) e^{4 \phi}
+ dz^2 e^{-4 \phi}
\end{equation}
The corresponding potential $A_{\mu}$ will be the same as for 2+1 dimensions.
Embedding the electric solution in 2+1 dimensions in eq.(49),
one can obtain a new rotating charged solution in 3+1 dimensions as,
$$
ds_{3+1}^2 =  \left( \frac{ r^{2}  L^4 \alpha^2 \gamma^2 -  \alpha^2  }
{  L^2} \right) dt^2 + 2 \left( \frac{ r^{2}  L^4 \alpha \gamma ( \alpha^{-1} 
+ \alpha \beta \gamma)
 -  \alpha^2 \beta   } {L^2} \right) dt d \varphi +$$
 \begin{equation}
\left( \frac{ r^{2}  L^4  ( \alpha^{-1} + \alpha \beta \gamma)^2
 -  \alpha^2 \beta^2  }
{  L^2} \right) d \varphi^2
+  r^{2m^2} L^2 ( dr^2 + dz^2 )
\end{equation} 
 The electromagnetic potential is given by,
\begin{equation}
\tilde{A_i} dx^i = \frac{\alpha Q}{2 c_1 m} 
\frac{r^{-m}}{(c_2 r^{-m} + c_1 r^{m})} ( dt + \beta d\varphi)
\end{equation}
Here, $L$ is a function of $r$ given by,
\begin{equation}
L = c_1 r^m + c_2 r^{-m}
\end{equation}
\newpage
Embedding the magnetic solution in 2+1 dimensions in eq.(52),
one can obtain a new rotating charged solution in 3+1 dimensions as,
$$
ds_{3+1}^2 =  \left( \frac{ - r^{2 + 2m^2} L^4 \alpha^2 + r^2 \alpha^2 \gamma^2 }
{ r^2 L^2} \right) dt^2 +
2\left( \frac{ - r^{2 + 2m^2} L^4 \alpha^2 \beta  
+ r^2 \alpha \gamma ( \alpha^{-1} + \alpha \beta \gamma) }
{ r^2 L^2} \right) dt d \varphi $$
\begin{equation} +
\left( \frac{ - r^{2 + 2m^2} L^4 \alpha^2 \beta^2  
+ r^2  ( \alpha^{-1} + \alpha \beta \gamma)^2 }
{ r^2 L^2} \right) d \varphi^2  + r^2 L^{ 2m^2} dr^2 + r^2 L^2 dz^2
 \end{equation}
The electromagnetic potential is given by,
\begin{equation}
\tilde{A_i} dx^i = \frac{-\alpha \gamma Q}{2 c_1 m} 
\frac{r^{-m}}{(c_2 r^{-m} + c_1 r^{m})} \left( dt + ( \alpha^{-2} \gamma^{-1} 
+\beta) d\varphi \right)
\end{equation}
Here, $L$ is a function of $r$ given by,
\begin{equation}
L = c_1 r^m + c_2 r^{-m}
\end{equation}

\section{Conclusions}
We have discussed
a new family of charged-dilaton solutions in 2+1
dimensions and charged solutions with cylindrical symmetry
in 3+1 dimensions.
A class of static charged solutions in 2+1 dimensions
are obtained by compactifying solutions with cylindrical
symmetry in 3+1 dimensions. These space-times contain
flat metric and naked singularities for certain values
of the parameters considered. We have 
applied ``T-duality'' transformations
to generate new rotating charged solutions in 2+1 dimensions. 
These rotating charged solutions lead to new
rotating charged solutions with cylindrical symmetry
once  embedded in 3+1 dimensions.

Since all the solutions presented here contain wide number of
parameters which may lead to space-times with differing physical
properties, it is of a worthy cause
to  analysis  these solutions in detail.
A particular interesting solution
to study further is the solution obtained from the Melvin
solution in 3+1 dimensions. It is well known that the
Melvin solution has direct generalizations to low-energy string theory
\cite{lowenergy}. It
has been shown that Melvin solution can be realized as the 
Kaluza-Klein reduction from flat higher dimensional space-time with non
trivial identifications\cite{dowker}.
Therefore, the dimensionally reduced Melvin solution
to 2+1 dimensions has direct relations to higher dimensional
space-times such as solutions of 11 dimensional M-theory.
Studying rotating  Melvin solution would be an interesting
exercise to further our understanding of higher dimensional 
physics.

It would be also interesting  to embed the solutions discussed here
in a supergravity theory arising from a low energy string theory
along the lines of supersymmetric solutions to
three dimensional heterotic string action considered
by Bakas {\it et.al.}\cite{bakas}.
We hope to address these
issues in the future.

\end{document}